\documentclass[letterpaper]{jpconf}
\usepackage{graphicx}
\usepackage{amsmath,amssymb,amsfonts,amsthm}
\newcommand{\pd}{\partial}
\newcommand{\under}[1]{_{#1}}
\numberwithin{equation}{section}
\begin{document}
\title{The Dunkl oscillator in three dimensions}

\author{Vincent X. Genest}
\address{Centre de Recherches Math\'ematiques, Universit\'e de Montr\'eal, Montr\'eal, Qu\'ebec, Canada}
\ead{vincent.genest@umontreal.ca}
\author{Luc Vinet}
\address{Centre de Recherches Math\'ematiques, Universit\'e de Montr\'eal, Montr\'eal, Qu\'ebec, Canada}
\ead{luc.vinet@umontreal.ca}
\author{Alexei Zhedanov}
\address{Donetsk Institute for Physics and Technology, Donetsk, Ukraine}
\ead{zhedanov@yahoo.com}

\begin{abstract}
The isotropic Dunkl oscillator model in three-dimensional Euclidean space is considered. The system is shown to be maximally superintegrable and its symmetries are obtained by the Schwinger construction using the raising/lowering operators of the dynamical $sl_{-1}(2)$ algebra of the one-dimensional Dunkl oscillator. The invariance algebra generated by the constants of motion, an extension of $\mathfrak{u}(3)$ with reflections, is called the Schwinger-Dunkl algebra $sd(3)$. The system is shown to admit separation of variables in Cartesian, polar (cylindrical) and spherical coordinates and the corresponding separated solutions are expressed in terms of generalized Hermite, Laguerre and Jacobi polynomials.
\end{abstract}

\section{Introduction}
This paper is concerned with the study of the isotropic Dunkl oscillator model in three-dimensional Euclidean space. This model, described by a Hamiltonian involving reflection operators, will be shown to be both maximally superintegrable and exactly solvable. The constants of motion will be obtained by the Schwinger construction using the $sl_{-1}(2)$ dynamical symmetry of the parabose oscillator in one dimension. The invariance algebra generated by the symmetries will be seen to be an extension of $\mathfrak{u}(3)$ by involutions and shall be called the Schwinger-Dunkl algebra $sd(3)$. The Schr\"odinger equation of the system will be seen to admit separation of variables in Cartesian, polar (cylindrical) and spherical coordinates. The corresponding separated solutions will be found and expressed in terms of generalized Hermite, Laguerre and Jacobi polynomials.
\subsection{Superintegrability}
Let us first recall the notion of superintegrability. A quantum system with $d$ degrees of freedom defined by a Hamiltonian $H$ is \emph{maximally superintegrable} if it admits $2d-1$ algebraically independent symmetry operators $S_{i}$ that commute with the Hamiltonian, i.e.
\begin{align*}
[H,S_i]=0,\qquad 1\leq i \leq  2d-1,
\end{align*}
where one of the symmetries is the Hamiltonian itself, e.g. $S_1=H$. A superintegrable system is said to be of order $\ell$ if $\ell$ is the maximal degree of the constants of motion $S_i$, excluding the Hamiltonian, in the momentum variables. The $\ell=1$ case is associated to symmetries of geometric nature and to Lie algebras whereas the $\ell=2$ case is typically associated to multiseparability of the Schr\"odinger equation and to quadratic invariance algebras.

A substantial amount of work has been done on superintegrable systems, motivated in part by their numerous applications, exact solvability and connections with the theory of special functions. In view of these properties, the search for new superintegrable models and their characterization is of significant interest in mathematical physics. For a recent review on superintegrable systems, one can consult \cite{Miller-2013-10}. 

\subsection{Dunkl oscillator models in the plane}
A series of novel superintegrable models in the plane with Hamiltonians involving reflection operators have been introduced recently \cite{Genest-2013-04,Genest-2013-09,Genest-2013-07}. The simplest of these systems, called the Dunkl oscillator in the plane, is defined by the following Hamiltonian \cite{Genest-2013-04}:
\begin{align}
\label{Dunkl-2D-H}
H=-\frac{1}{2}\Big[\mathcal{D}_{1}^2+\mathcal{D}_2^{2}\Big]+\frac{1}{2}\Big[x_1^2+x_2^2\Big],
\end{align}
where $\mathcal{D}_{i}$ stands for the Dunkl derivative \cite{Dunkl-1989-01}
\begin{align}
\label{Dunkl-D}
\mathcal{D}_i=\pd_{i}+\frac{\mu_i}{x_i}\big(1-R_{i}\big),\qquad i=1,2,
\end{align}
with $\pd_{i}=\frac{\pd}{\pd x_i}$ and where $R_i$ is the reflection operator with respect to the $x_i=0$ axis, i.e
\begin{align*}
R_1 f(x_1,x_2)=f(-x_1,x_2),\qquad R_2 f(x_1,x_2)=f(x_1,-x_2).
\end{align*}
The Hamiltonian \eqref{Dunkl-2D-H} can obviously be written as
\begin{align*}
H=H_1+H_2
\end{align*}
where $H_i$, $i=1,2$, is the one-dimensional Dunkl oscillator Hamiltonian
\begin{align}
\label{1D-Dunkl}
H_i=-\frac{1}{2}\mathcal{D}_i^2+\frac{1}{2}x_i^2.
\end{align}
In \cite{Genest-2013-04}, the Hamiltonian \eqref{Dunkl-2D-H} was shown to be maximally superintegrable and its two independent constants of motion, obtained by the Schwinger construction, were seen to generate a $\mathfrak{u}(2)$ algebra extended with involutions. This algebra was called the Schwinger-Dunkl algebra $sd(2)$. It was further shown that the Schr\"odinger equation associated to \eqref{Dunkl-2D-H} admits separation of variables in both Cartesian and polar coordinates and the separated wavefunctions were given explicitly. Furthermore, the overlap coefficients between the separated wavefunctions were expressed in terms of the dual $-1$ Hahn polynomials through a correspondence with the Clebsch-Gordan problem of $sl_{-1}(2)$, a $q\rightarrow -1$ limit of the quantum algebra $sl_{q}(2)$ (see Section II for the definition of $sl_{-1}(2)$ and \cite{Tsujimoto-2011-10} for background). In \cite{Genest-2013-09}, the representation theory of the Schwinger-Dunkl algebra $sd(2)$ was examined. The results obtained in this paper further strengthened the idea that Dunkl oscillator models are showcases for $-1$ orthogonal polynomials (see \cite{Genest-2013-02-1,Tsujimoto-2012-03}). 

Generalizations of the Dunkl oscillator in the plane \eqref{Dunkl-2D-H} were considered in \cite{Genest-2013-07}, where the singular and the $2:1$ anisotropic Dunkl oscillators were investigated. The two systems were shown to be superintegrable, their constants of motion were constructed and their (quadratic) invariance algebra was given. It was also shown that in some cases these models exhibit multiseparability. Here we pursue our investigations on Dunkl oscillator models by considering the three-dimensional version of the system \eqref{Dunkl-2D-H} in Euclidean space.
\subsection{The three-dimensional Dunkl oscillator}
The Dunkl oscillator in three-dimensional Euclidean space is defined by the Hamiltonian
\begin{align}
\label{3D-Dunkl-H}
\mathcal{H}=-\frac{1}{2}\Big[\mathcal{D}_1^2+\mathcal{D}_2^2+\mathcal{D}_3^2\Big]+\frac{1}{2}\Big[x_1^2+x_2^2+x_3^2\Big],
\end{align}
where $\mathcal{D}_i$ is the Dunkl derivative given by \eqref{Dunkl-D}. An elementary calculation shows that the square of the Dunkl derivative has the expression
\begin{align*}
\mathcal{D}_i^2=\pd_{i}^2+\frac{2\mu_i}{x_i}\,\pd_{i}-\frac{\mu_i}{x_i^2}(1-R_i).
\end{align*}
As is directly seen, the Hamiltonian \eqref{3D-Dunkl-H} corresponds to the combination of three independent (commuting) one-dimensional Dunkl oscillators with Hamiltonians $H_i$ given by \eqref{1D-Dunkl}. It directly follows from the formulas \eqref{Dunkl-D} and \eqref{3D-Dunkl-H} that when $\mu_i=0$, $i=1,2,3$, the Dunkl oscillator Hamiltonian \eqref{3D-Dunkl-H} reduces to that of the standard isotropic harmonic oscillator in three dimensions. We note that the term ``isotropic'' here refers to the fact that the three independent one-dimensional oscillator entering the total Hamiltonian \eqref{3D-Dunkl-H} have the same frequency. To obtain a 3D Dunkl oscillator isotropic in the sense of being $O(3)$-invariant, one must also take $\mu_1=\mu_2=\mu_3$.
\subsection{Outline}
The paper is organized as follows. In section II, the $sl_{-1}(2)$ dynamical symmetry and the Schwinger construction are used to obtain the symmetries of the total Hamiltonian \eqref{3D-Dunkl-H}, thus establishing its superintegrability. The commutation relations satisfied by the constants of motion, which define the Schwinger-Dunkl algebra $sd(3)$, are also exhibited in two different bases. In section III, the separated solutions of the Schr\"odinger equation associated to \eqref{3D-Dunkl-H} are given in Cartesian, polar and spherical coordinates. For each coordinate system, the symmetries responsible for the separation of variables are given. A short conclusion follows.
\section{Superintegrability}
In this section, the $sl_{-1}(2)$ dynamical algebra of the three-dimensional Dunkl oscillator is made explicit and is exploited to obtain the spectrum of the Hamiltonian. Following the Schwinger construction, the constants of motion are constructed using the raising/lowering operators of the $sl_{-1}(2)$ algebra of the one-dimensional constituents. The Schwinger-Dunkl algebra $sd(3)$ formed by the symmetries is seen to correspond to a deformation of the Lie algebra $\mathfrak{u}(3)$ by involutions and the defining relations of $sd(3)$ are given in two different bases.
\subsection{Dynamical algebra and spectrum}
The three-dimensional Dunkl oscillator Hamiltonian \eqref{3D-Dunkl-H} possesses an $sl_{-1}(2)$ dynamical symmetry inherited from the one of its one-dimensional constituents $H_i$. This can be seen as follows. Consider the following operators: 
\begin{align}
\label{Creation/Annihilation}
A_{\pm}^{(i)}=\frac{1}{\sqrt{2}}\Big(x_i\mp\mathcal{D}_i\Big),
\end{align}
and define $A_0^{(i)}=H_i$ with $H_i$ given by \eqref{1D-Dunkl}. It is easily checked that these operators, together with the reflection operator $R_i$, satisfy the defining relations of the $sl_{-1}(2)$ algebra which are of the form \cite{Tsujimoto-2011-10}
\begin{align}
\label{sl}
[A_0^{(i)},A_{\pm}^{(i)}]=\pm A_{\pm}^{(i)},\quad [A_0^{(i)},R_i]=0,\quad \{A_{+}^{(i)},A_{-}^{(i)}\}=2A_0^{(i)},\quad \{A_{\pm}^{(i)},R_{i}\}=0,
\end{align}
where $[x,y]=xy-yx$ and $\{x,y\}=xy+yx$. As is seen from the above commutation relations, the  operators $A_{\pm}^{(i)}$ act as raising/lowering operators for the one-dimensional Hamiltonians $A_{0}^{(i)}=H_i$. In the realization \eqref{Creation/Annihilation}, the $sl_{-1}(2)$ Casimir operator 
\begin{align*}
Q^{(i)}=A_{+}^{(i)}A_{-}^{(i)}R_i-A_0^{(i)}R_i+R_i/2,
\end{align*}
is a multiple of the identity $Q^{(i)}=-\mu_i$. It thus follows that for $\mu_i>-1/2$, the operators $A^{(i)}_{\pm}$, $A_0^{(i)}$ and $R_i$ realize the positive-discrete series of representations of $sl_{-1}(2)$ \cite{Tsujimoto-2011-10}. In these representations, the spectrum $\mathcal{E}^{(i)}$ of the operators $A_{0}^{(i)}$ is given by
\begin{align*}
\mathcal{E}_{n_i}^{(i)}=n_i+\mu_i+1/2,
\end{align*}
where $n_i$ is a non-negative integer. Given that the full Hamiltonian \eqref{3D-Dunkl-H} of the three-dimensional Dunkl oscillator is of the form $\mathcal{H}=H_1+H_2+H_3$, it follows that its energy eigenvalues $\mathcal{E}$ have the expression
\begin{align}
\mathcal{E}_{N}=N+\mu_1+\mu_2+\mu_3+3/2,
\end{align}
where $N$ is a non-negative integer. Since the integer $N$ can be written as $N=n_1+n_2+n_3$, the spectrum of $\mathcal{H}$ has degeneracy
\begin{align*}
g_{N}=\sum_{n_3=0}^{N}(N-n_3+1)=(N+1)(N+2)/2,
\end{align*}
at energy level $\mathcal{E}_{N}$. Hence the three-dimensional Dunkl oscillator exhibits the same degeneracy in its spectrum as the standard three-dimensional isotropic oscillator.

The coproduct of $sl_{-1}(2)$ (see \cite{Tsujimoto-2011-10}) can be used to construct the $sl_{-1}(2)$ dynamical algebra of the total Hamiltonian $\mathcal{H}$. Indeed, upon defining the operators
\begin{align*}
\mathcal{A}_{\pm}=A_{\pm}^{(1)}R_{2}R_{3}+A_{\pm}^{(2)}R_{3}+A_{\pm}^{(3)}, \quad \mathcal{R}=R_{1}R_{2}R_{3},\qquad \mathcal{A}_0=\mathcal{H},
\end{align*}
it is directly checked that one has
\begin{align*}
[\mathcal{A}_0,\mathcal{A}_{\pm}]=\pm \mathcal{A}_{\pm},\quad [\mathcal{A}_0,\mathcal{R}]=0,\quad \{\mathcal{A}_{+},\mathcal{A}_{-}\}=2\mathcal{A}_0,\quad \{\mathcal{A}_{\pm},\mathcal{R}\}=0.
\end{align*}
Hence the operators $\mathcal{A}_{\pm}$ act as the raising/lowering operators for the full Hamiltonian $\mathcal{H}$ of the three-dimensional Dunkl oscillator.
\subsection{Constants of motion and the Schwinger-Dunkl algebra $sd(3)$}
Given that $\mathcal{H}=H_1+H_2+H_3$, it is clear that combining a raising operator for $H_i$ with a lowering operator for $H_j$ when $i\neq j$ will result in an operator preserving the eigenspace associated to a given energy $\mathcal{E}_{N}$, thus producing a constant of motion of the total Hamiltonian $\mathcal{H}$. Moreover, it is obvious that the one-dimensional components $H_i$ commute with the total Hamiltonian $\mathcal{H}$ and hence these one-dimensional Hamiltonians can be considered as constants of the motion. Furthermore, it is observed that the total Hamiltonian commutes with the reflection operators
\begin{align*}
[\mathcal{H},R_i]=0,\qquad i=1,2,3,
\end{align*}
so that the reflections can be considered as symmetries. Following the Schwinger construction of $\mathfrak{u}(3)$ with standard creation/annihilation operators, we define
\begin{gather*}
J_{1}=\frac{1}{i}\big(A_{+}^{(2)}A_{-}^{(3)}-A_{-}^{(2)}A_{+}^{(3)}\big),\quad
J_{2}=\frac{1}{i}\big(A_{+}^{(3)}A_{-}^{(1)}-A_{-}^{(3)}A_{+}^{(1)}\big),
\\
J_{3}=\frac{1}{i}\big(A_{+}^{(1)}A_{-}^{(2)}-A_{-}^{(1)}A_{+}^{(2)}\big).
\end{gather*}
The operators $J_i$, $i=1,2,3$, can be interpreted as Dunkl ``rotation'' generators since in terms of Dunkl derivatives, they read
\begin{align}
\label{Rotations}
J_1=\frac{1}{i}(x_2 \mathcal{D}_3-x_3 \mathcal{D}_2),\quad J_2=\frac{1}{i}(x_3 \mathcal{D}_1-x_1 \mathcal{D}_3),\quad J_3=\frac{1}{i}(x_1 \mathcal{D}_2-x_2 \mathcal{D}_1).
\end{align}
We also introduce the operators
\begin{align*}
K_1=\big(A_{+}^{(2)}A_{-}^{(3)}+A_{-}^{(2)}A_{+}^{(3)}\big),\;
K_2=\big(A_{+}^{(3)}A_{-}^{(1)}+A_{-}^{(3)}A_{+}^{(1)}\big),\;
K_3=\big(A_{+}^{(1)}A_{-}^{(2)}+A_{-}^{(1)}A_{+}^{(2)}\big),
\end{align*}
which in coordinates have the expression
\begin{align}
\label{Boosts}
K_1=(x_2x_3-\mathcal{D}_2\mathcal{D}_3),\quad K_2=(x_3x_1-\mathcal{D}_3\mathcal{D}_1),\quad K_3=(x_1x_2-\mathcal{D}_1\mathcal{D}_2).
\end{align}
It is directly checked that the operators $J_i$, $K_i$, $i=1,2,3$, are symmetries of the total Dunkl oscillator Hamiltonian \eqref{3D-Dunkl-H}, that is
\begin{align*}
[J_i,\mathcal{H}]=[K_i,\mathcal{H}]=0,
\end{align*}
for $i=1,2,3$. Upon writing
\begin{align*}
L_1=H_1/2,\quad L_2=H_2/2,\quad L_3=H_3/2,
\end{align*}
which satisfy $\mathcal{H}=2L_1+2L_2+2L_3$, a direct computation shows that the non-zero commutation relations between the symmetries are given by
\begin{subequations}
\label{sd(3)}
\begin{align}
[J_j,J_k]=i\epsilon_{jk\ell}J_{\ell}(1+2\mu_{\ell}R_{\ell}),\qquad 
[K_{j},K_{k}]=-i\epsilon_{jk\ell}J_{\ell}(1+2\mu_{\ell}R_{\ell}),
\end{align}
where $\epsilon_{ijk}$ stands for the totally antisymmetric tensor with summation over repeated indices understood and
\begin{gather}
[J_j,K_k]=-i\epsilon_{jk\ell}K_{\ell}(1+2\mu_{\ell}R_{\ell}),\quad 
[J_j, K_{j}]=-i\epsilon_{jk\ell}L_{k}(1+2\mu_{\ell}R_{\ell}),
\end{gather}
for $j\neq k$. One also has
\begin{gather}
[J_j,L_k]=ig_{jk}K_j/2,\quad [K_j,L_{k}]=-ig_{jk}J_{j}/2,
\end{gather}
where $g_{ij}$ are the elements of a $3\times 3$ antisymmetric matrix with $g_{12}=g_{23}=1$, $g_{13}=-1$. The commutation relations involving the reflection operators are of the form
\begin{gather}
[L_i,R_{i}]=[L_{j},R_{k}]=[J_i,R_i]=[K_i,R_i]=0,
\\
\{J_{j},R_{k}\}=\{K_{j},R_{k}\}=0,
\end{gather}
\end{subequations}
where $j\neq k$. Hence it follows that the three-dimensional Dunkl oscillator model is maximally superintegrable. The invariance algebra is defined by the commutation and anticommutation relations \eqref{sd(3)} and we shall refer to this algebra as the Schwinger-Dunkl algebra $sd(3)$. It is clear from the defining relations \eqref{sd(3)} that the algebra $sd(3)$ corresponds to a deformation of the $\mathfrak{u}(3)$ Lie algebra by involutions; the central element here is of course the total Hamiltonian $\mathcal{H}=H_1+H_2+H_3$. If one takes $\mu_1=\mu_2=\mu_3=0$ in the commutation relations \eqref{sd(3)}, one recovers the $\mathfrak{u}(3)$ symmetry algebra of the standard isotropic harmonic oscillator in three dimensions realized by the standard creation/annihilation operators.
\subsection{An alternative presentation of $sd(3)$}
The Schwinger-Dunkl algebra $sd(3)$ obtained here can be seen as a ``rank two'' version of the Schwinger-Dunkl algebra $sd(2)$ which has appeared in \cite{Genest-2013-04} as the symmetry algebra of the Dunkl oscillator in the plane. It is possible to present another basis for the symmetries of the three-dimensional Dunkl oscillator in which the $sd(2)$ algebra explicitly appears as a subalgebra of $sd(3)$. In order to define this basis, it is convenient to introduce the standard $3\times 3$ Gell-Mann matrices \cite{Arfken-2012} denoted by $\Lambda_i$, $i=1,\ldots,8$, and obeying the $\mathfrak{su}(3)$ commutation relations
\begin{align*}
[\Lambda_i,\Lambda_j]=i f^{ijk}\Lambda_{k},
\end{align*}
with $f^{123}=2$, $f^{458}=f^{678}=\sqrt{3}$ and
\begin{align*}
f^{147}=f^{165}=f^{246}=f^{257}=f^{345}=f^{376}=1.
\end{align*}
The symmetries of the three-dimensional Dunkl oscillator can be expressed in terms of these matrices as follows. One takes
\begin{align}
M_{j}=(A_{+}^{(1)},A_{+}^{(2)},A_{+}^{(3)})\;\Lambda_j\;(A_{-}^{(1)},A_{-}^{(2)},A_{-}^{(3)})^{t},
\end{align}
for $j=1,2,4,5,6,7$ and also
\begin{align*}
M_3&=\frac{1}{4}\left(\{A_{+}^{(1)},A_{-}^{(1)}\}-\{A_{+}^{(2)}A_{-}^{(2)}\}\right),
\\
M_8&=\frac{1}{4\sqrt{3}}\Big(\{A_{+}^{(1)},A_{-}^{(1)}\}+\{A_{+}^{(2)},A_{-}^{(2)}\}-2\{A_{+}^{(3)},A_{-}^{(3)}\}\Big).
\end{align*}
It is straightforward to verify that the operators $M_i$, $i=1,\ldots,8$ commute with the Hamiltonian $\mathcal{H}$ of the three-dimensional Dunkl oscillator. Using the commutation relations \eqref{sl} as well as the extra relation
\begin{align*}
[A_{-}^{(i)},A_{+}^{(j)}]=\delta_{ij}(1+2\mu_i R_{i}),
\end{align*}
the $sd(3)$ commutation relations expressed in the basis of the symmetries $M_i$ can easily be obtained. Consider the constants of motion
\begin{align*}
M_1=\frac{1}{2}\left(A_{+}^{(1)}A_{-}^{(2)}+A_{-}^{(1)}A_{+}^{(2)}\right),\quad M_2=\frac{1}{2i}\left(A_{+}^{(1)}A_{-}^{(2)}-A_{-}^{(1)}A_{+}^{(2)}\right),
\end{align*}
as well as $M_3$. These symmetry operators satisfy the commutation relations of the Schwinger-Dunkl algebra $sd(2)$
\begin{subequations}
\label{sd(2)}
\begin{gather}
[M_2,M_3]=iM_1,\quad [M_3, M_1]=i M_2,
\\
[M_1,M_2]=i\Big(M_3+M_3(\mu_1R_1+\mu_2R_2)-\frac{1}{3}\left(\mathcal{H}+\sqrt{3}M_8\right)(\mu_1R_1-\mu_2 R_2)\Big),
\\
\{M_1,R_{i}\}=0,\quad \{M_2,R_{i}\}=0,\quad [M_3,R_i]=0,
\end{gather}
for $i=1,2$. In the subalgebra \eqref{sd(2)}, the operator $\left(\mathcal{H}+\sqrt{3}M_8\right)$ is central and corresponds to the Hamiltonian of the Dunkl oscillator in the plane
\begin{align*}
\frac{2}{3}\left(\mathcal{H}+\sqrt{3}M_8\right)=H_{1}+H_{2}.
\end{align*}
\end{subequations}
\section{Separated Solutions: Cartesian, cylindrical and spherical coordinates}
In this section, the exact solutions of the time-independent Schr\"odinger equation
\begin{align}
\label{Sch}
\mathcal{H}\Psi=\mathcal{E}\Psi,
\end{align}
associated to the three-dimensional Dunkl oscillator Hamiltonian \eqref{3D-Dunkl-H} are obtained in Cartesian, polar (cylindrical) and spherical coordinates. The operators responsible for the separation of variables in each of these coordinate systems are given explicitly.
\subsection{Cartesian coordinates}
Since $\mathcal{H}=H_1+H_2+H_3,$ where $H_i$, $i=1,2,3$, are the one-dimensional Dunkl oscillator Hamiltonians
\begin{align*}
H_i=-\frac{1}{2}\mathcal{D}_i^2+\frac{1}{2}x_i^2,
\end{align*}
it is obvious that the Schr\"odinger equation \eqref{Sch} admits separation of variable in Cartesian coordinates $\{x_1,x_2,x_3\}$. In this coordinate system, the separated solutions are of the form
\begin{align*}
\Psi(x_1,x_2,x_3)=\psi(x_1)\psi(x_2)\psi(x_3),
\end{align*}
where $\psi(x_i)$ are solutions of the one-dimensional Schr\"odinger equation 
\begin{align}
\label{Sch-2}
\left[-\frac{1}{2}\mathcal{D}_i^2+\frac{1}{2}x_i^2\right]\psi(x_i)=\mathcal{E}^{(i)}\psi(x_i).
\end{align}
The regular solutions of \eqref{Sch-2} are well known \cite{Genest-2013-04,Mukunda-1980-10}. To obtain these solutions, one uses the fact that the reflection operator $R_i$ commutes with the one-dimensional Hamiltonian $H_i$, which allows to diagonalize both operators simultaneously. For the one-dimensional problem, the two sectors corresponding to the possible eigenvalues $s_i=\pm 1$ of the reflection operator $R_i$ can be recombined to give the following expression for the wavefunctions:
\begin{align}
\label{1D-Wave}
\psi_{n_i}(x_i)=e^{-x_i^2/2}H_{n_i}^{\mu_i}(x_i),
\end{align}
where $n_i$ is a non-negative integer. The corresponding energy eigenvalues are
$$
\mathcal{E}^{(i)}_{n_i}=n_i+\mu_i+1/2,
$$
and $H_{n}^{\mu}(x)$ stands for the generalized Hermite polynomials \cite{Chihara-2011}
\begin{align}
\label{Gen-H}
H_{2m+p}^{\mu}(x)=(-1)^{n}\sqrt{\frac{n!}{\Gamma(m+p+\mu+1/2)}}\,x^{p}\,L_{m}^{(\mu-1/2+p)}(x^2),
\end{align}
with $p\in \{0,1\}$. In \eqref{Gen-H}, $L_{n}^{(\alpha)}(x)$ are the standard Laguerre polynomials \cite{Koekoek-2010} and $\Gamma(x)$ is the classical Gamma function \cite{Arfken-2012}. The wavefunctions \eqref{1D-Wave} satisfy
\begin{align*}
R_{i}\psi_{n_i}(x_i)=(-1)^{n_i}\psi_{n_i}(x_i),
\end{align*}
so that the eigenvalue $s_i$ of the reflection operator $R_i$ is given by the parity of $n_{i}$. Using the orthogonality relation of the Laguerre polynomials, one finds that the wavefunctions \eqref{1D-Wave} satisfy the orthogonality condition
\begin{align*}
\int_{-\infty}^{\infty}\psi_{n_i}(x_i)\psi_{n_i'}(x_i)|x_i|^{2\mu_i}\,\mathrm{d}x_i=\delta_{n_i,n_i'}.
\end{align*}
In Cartesian coordinates, the separated solution of the Schr\"odinger equation associated to the three-dimensional Dunkl oscillator Hamiltonian \eqref{3D-Dunkl-H} are thus given by
\begin{align}
\label{Cartesian-Wave}
\Psi_{n_1,n_2,n_3}(x_1,x_2,x_3)=e^{-(x_1^2+x_2^2+x_3^2)/2}H_{n_1}^{\mu_1}(x_1)H_{n_2}^{\mu_2}(x_2)H_{n_3}^{\mu_3}(x_3),
\end{align}
and the corresponding energy $\mathcal{E}$ is
\begin{align}
\label{Cartesian-Energy}
\mathcal{E}=n_1+n_2+n_3+\mu_1+\mu_2+\mu_3+3/2,
\end{align}
where $n_i$, $i=1,2,3$, are non-negative integers. It is directly seen from \eqref{Gen-H}, \eqref{Cartesian-Wave} and \eqref{Cartesian-Energy} that if one takes $\mu_i=0$, the solutions and the spectrum of the isotropic three-dimensional oscillator in Cartesian coordinates are recovered.
\subsection{Cylindrical coordinates}
In cylindrical coordinates
\begin{align*}
x_1=\rho \cos \varphi,\quad x_2=\rho \sin \varphi,\quad x_3=z.
\end{align*}
the Hamiltonian \eqref{3D-Dunkl-H} of the three-dimensional Dunkl oscillator reads
\begin{align*}
\mathcal{H}=\mathcal{A}_{\rho}+\frac{1}{\rho^2}\mathcal{B}_{\varphi}+\mathcal{C}_{z},
\end{align*}
where
\begin{subequations}
\begin{align}
\mathcal{A}_{\rho}&=-\frac{1}{2}\left[\pd_{\rho}^2+\frac{1}{\rho}\pd_{\rho}\right]-\frac{(\mu_1+\mu_2)}{\rho}\pd_{\rho}+\frac{1}{2}\rho^2,
\\
\label{B}
\mathcal{B}_{\varphi}&=-\frac{1}{2}\pd_{\varphi}^2+(\mu_1\tan \varphi-\mu_2\cot \varphi)\pd_{\varphi}+\frac{\mu_1}{2\cos^2\varphi}(1-R_1)+\frac{\mu_2}{2\sin^2\varphi}(1-R_2),
\\
\mathcal{C}_z&=-\frac{1}{2}\pd_{z}^2-\frac{\mu_3}{z}\pd_{z}+\frac{1}{2}z^2+\frac{\mu_3}{2z^2}(1-R_3).
\end{align}
\end{subequations}
The reflection operators are easily seen to have the action
\begin{align*}
R_1 f(\rho,\varphi,z)=f(\rho,\pi-\phi,z),\quad R_2 f(\rho,\varphi,z)=f(\rho,-\varphi,z),\quad R_3 f(\rho,\varphi,z)=f(\rho,\varphi,-z).
\end{align*}
Upon taking $\Psi(\rho,\varphi,z)=P(\rho)\Phi(\varphi)\psi(z)$, one finds that \eqref{Sch} is equivalent to the system of ordinary equations
\begin{subequations}
\begin{align}
\label{a}
&\mathcal{A}_{\rho}P(\rho)-\widetilde{\mathcal{E}}P(\rho)+\frac{k^2}{2\rho^2}P(\rho)=0,
\\
\label{b}
&\mathcal{B}_{\varphi}\Phi(\varphi)-\frac{k^2}{2}\Phi(\varphi)=0,
\\
\label{c}
&\mathcal{C}_{z}\psi(z)=\mathcal{E}^{(3)}\psi(z),
\end{align}
\end{subequations}
where $\mathcal{E}^{(3)}$, $k^2/2$ are the separation constants and where $\widetilde{\mathcal{E}}=\mathcal{E}-\mathcal{E}^{(3)}$. The solutions to the equation are given by \eqref{1D-Wave} and \eqref{Gen-H} with $\mathcal{E}^{(3)}=n_3+\mu_3+1/2$. The solutions to \eqref{a} and \eqref{b} have been obtained in \cite{Genest-2013-04}. For the angular part, the solutions are labeled by the eigenvalues $s_1, s_2$ with $s_i=\pm 1$ of the reflection operators $R_1$, $R_2$ and read
\begin{align}
\label{Angular}
\Phi^{(s_1,s_2)}_{m}(\varphi)=\eta_{m}\cos^{e_1}\varphi \sin^{e_2}\varphi\,P_{m-e_1/2-e_2/2}^{(\mu_2+e_2-1/2,\mu_1+e_1-1/2)}(\cos 2\varphi),
\end{align}
where $P_{n}^{(\alpha,\beta)}(x)$ are the classical Jacobi polynomials \cite{Koekoek-2010}, $\eta_{m}$ is a normalization factor and where $e_1$, $e_2$ are the indicator functions for the eigenvalues of the reflections $R_1$ and $R_2$, i.e.:
\begin{align*}
e_i=
\begin{cases}
0, & \text{if $s_i=1$},\\
1, & \text{if $s_i=-1$},
\end{cases}
\end{align*}
for $i=1,2$. When $s_1s_2=-1$, $m$ is a positive half-integer whereas when $s_1s_2=1$, $m$ is a non-negative integer; note also that for $m=0$, only the $s_1=s_2=1$ state exists. In all parity cases, the separation constant takes the value
\begin{align*}
k^2=4m(m+\mu_1+\mu_2).
\end{align*}
If one takes
\begin{align*}
\eta_{m}=\left[\frac{(2m+\mu_1+\mu_2)\Gamma(m+\mu_1+\mu_2+\frac{e_1}{2}+\frac{e_2}{2})(m-\frac{e_1}{2}-\frac{e_2}{2})!}{2\,\Gamma(m+\mu_1+\frac{e_1}{2}-\frac{e_2}{2}+1/2)\,\Gamma(m+\mu_2+\frac{e_2}{2}-\frac{e_1}{2}+1/2)}\right]^{1/2},
\end{align*}
as the normalization factor, the angular part of the separated wavefunction satisfy the orthogonality relation
\begin{align*}
\int_{0}^{2\pi}\Phi_{m}^{(s_1,s_2)}\Phi_{m'}^{(s_1',s_2')}\,\rvert\cos\phi\rvert^{2\mu_1}\,\rvert\sin\phi\rvert^{2\mu_2}\;\mathrm{d}\phi=\delta_{m,m'}\delta_{s_1,s_1'}\delta_{s_2,s_2'},
\end{align*}
which can be deduced from the orthogonality relation satisfied by the Jacobi polynomials \cite{Koekoek-2010}.
The radial solutions have the expression
\begin{align}
\label{Radial}
P_{n_{\rho}}(\rho)=\left[\frac{2\,n_{\rho}!}{\Gamma(n_{\rho}+2m+\mu_1+\mu_2+1)}\right]^{1/2}e^{-\rho^2/2}\rho^{2m}L_{n_{\rho}}^{(2m+\mu_1+\mu_2)}(\rho^2),
\end{align}
where $n_{\rho}$ is a non-negative integer and where $L_{n}^{(\alpha)}(x)$ are the Laguerre polynomials. They satisfy the orthogonality relation
\begin{align*}
\int_{0}^{\infty}P_{n_{\rho}}(\rho)P_{n_{\rho}'}(\rho)\,\rho^{2\mu_1+2\mu_2+1}\,\mathrm{d}\rho=\delta_{n_{\rho},n_{\rho}'}.
\end{align*}
The separated wavefunctions of the three-dimensional Dunkl oscillator in cylindrical coordinates are thus given by
\begin{align*}
\Psi_{n_{\rho},m,n_{z}}(\rho,\varphi,z)=P_{n_{\rho}}(\rho)\Phi_{m}^{(s_1,s_2)}(\varphi)\psi_{n_z}(z)
\end{align*}
where $P_{n_{\rho}}(\rho)$, $\Phi_{m}^{(s_1,s_2)}(\varphi)$ and $\psi_{n_z}(z)$ are given by \eqref{Radial}, \eqref{Angular} and \eqref{1D-Wave}, respectively. The energy $\mathcal{E}$ is expressed as
\begin{align*}
\mathcal{E}=2n_{\rho}+2m+n_z+\mu_1+\mu_2+\mu_3+3/2,
\end{align*}
where $n_{\rho}$, $n_{z}$ are non-negative integers and where $m$ is a non-negative integer when $s_1s_2=1$ or a positive half-integer when $s_1s_2=-1$. In the cylindrical basis, the operators ${\mathcal{C}}_{z}$ and $\mathcal{B}_{\phi}$ are diagonal with eigenvalues $\mathcal{E}^{(3)}$ and $k^2/2$. A direct computation shows that one has
\begin{align*}
J_3^2=2\mathcal{B}_{\varphi}+2\mu_1\mu_2(1-R_1R_2),
\end{align*}
where $J_3$ is the symmetry given in \eqref{Rotations}. It thus follows that $J_3$ and $H_3$ are the symmetries responsible for the separation of variables in cylindrical coordinates.
\subsection{Spherical coordinates}
In spherical coordinates
\begin{align*}
x_1=r\cos \phi \sin \theta,\quad x_2=r\sin \phi \sin \theta,\quad x_3=r \cos \theta,
\end{align*}
the Hamiltonian \eqref{3D-Dunkl-H} of the three-dimensional Dunkl oscillator takes the form
\begin{align*}
\mathcal{H}=\mathcal{M}_{r}+\frac{1}{r^2}\mathcal{N}_{\theta}+\frac{1}{r^2\sin^2\theta}\mathcal{B}_{\phi},
\end{align*}
where $\mathcal{B}_{\phi}$ is given by \eqref{B} and where 
\begin{subequations}
\begin{align}
\mathcal{M}_{r}&=-\frac{1}{2}\pd_{r}^2-\frac{(1+\mu_1+\mu_2+\mu_3)}{r}\pd_{r}+\frac{1}{2}r^2,
\\
\label{Azi}
\mathcal{N}_{\theta}&=-\frac{1}{2}\pd_{\theta}^2+(\mu_3\tan\theta-(1/2+\mu_1+\mu_2)\cot \theta)\pd_{\theta}+\frac{\mu_3}{2\cos^2\theta}(1-R_3).
\end{align}
\end{subequations}
The reflection operators have the action
\begin{align*}
R_1 f(r,\theta,\phi)=f(r,\theta,\pi-\phi),\quad R_2 f(r,\theta,\phi)=f(r,\theta,-\phi),\quad R_3 f(r,\theta,\phi)=f(r,\pi-\theta,\phi).
\end{align*}
Upon taking $\Psi(r,\theta,\phi)=R(r)\Theta(\theta)\Phi(\phi)$ in the Schr\"odinger equation \eqref{Sch}, one finds the following system of ordinary differential equations
\begin{subequations}
\begin{align}
&\left[\mathcal{M}_r+\left(\frac{q^2}{2r^2}-\mathcal{E}\right)\right]R(r)=0,
\\
&\left[\mathcal{N}_{\theta}+\left(\frac{k^2}{2\sin^2\theta}-\frac{q^2}{2}\right)\right]\Theta(\theta)=0,
\\
\label{Third}
&\left[\mathcal{B}_{\phi}-\frac{k^2}{2}\right]\Phi(\phi)=0,
\end{align}
\end{subequations}
where $k^2/2$ and $q^2/2$ are the separation constants. It is directly seen that the azimuthal solution $\Phi(\phi)$ to \eqref{Third} is given by \eqref{Angular} with value of the separation constant $k^2=4m(m+\mu_1+\mu_2)$. The zenithal solutions $\Theta(\theta)$ are labeled by the eigenvalue $s_3=\pm 1$ of the reflection operator $R_3$. One has
\begin{align}
\label{Zenithal}
\Theta_{\ell}^{(s_3)}(\theta)=\iota_{\ell}\cos^{e_3}\theta\sin^{2m}\theta\,P_{\ell-e_3/2}^{(2m+\mu_1+\mu_2,\mu_3+e_3-1/2)}(\cos 2\theta),
\end{align}
where the value of the separation constant is $q^2=4(\ell+m)(\ell+m+\mu_1+\mu_2+\mu_3+1/2)$. When $s_3=1$, $\ell$ is a non-negative integer whereas $\ell$ is a positive half-integer when $s_3=-1$. The normalization constant has the expression
\begin{align*}
\iota_{\ell}=\left[\frac{(2\ell+2m+\mu_1+\mu_2+\mu_3+1/2)\Gamma(\ell+2m+\mu_1+\mu_2+\mu_3+1/2+e_3/2)(\ell-e_3/2)!}{\Gamma(\ell+2m+\mu_1+\mu_2+1-e_3/2)\Gamma(\ell+\mu_3+1/2+e_3/2)}\right]^{1/2}.
\end{align*}
The radial solutions are given by
\begin{align}
\label{Radial-R}
R_{n_r}(r)=\left[\frac{2n_r!}{\Gamma(n_r+\alpha+1)}\right]^{1/2}e^{-r^2/2}r^{2(\ell+m)}L_{n_r}^{(\alpha)}(r^2),
\end{align}
with $\alpha=2(\ell+m)+\mu_1+\mu_2+\mu_3+1/2$. The separated wavefunctions of the three-dimensional Dunkl oscillator in spherical coordinates thus read
\begin{align}
\Psi_{n_r,\ell,m}^{(s_1,s_2,s_3)}(r,\theta,\phi)=R_{n_r}(r)\,\Phi_{m}^{(s_1,s_2)}(\phi)\,\Theta_{\ell}^{(s_3)}(\theta),
\end{align}
where $R(r)$, $\Theta(\theta)$ and $\Phi(\phi)$ are respectively given by \eqref{Radial-R}, \eqref{Zenithal} and \eqref{Angular} and correspond to the total energy
\begin{align}
\mathcal{E}=2(n_r+\ell+m)+\mu_1+\mu_2+\mu_3+3/2.
\end{align}
These wavefunctions are eigenfunctions of the reflection operators $R_i$ with eigenvalues $s_i=\pm 1$ for $i=1,2,3$. When $s_3=1$, the quantum number $\ell$ takes non-negative integer values and when $s_3=-1$, the number $\ell$ takes positive half-integer values. Similarly, when $s_1s_2=1$, the quantum number $m$ is a non-negative integer and when $s_1s_2=-1$, $m$ is a positive half-integer. The wavefunctions satisfy the orthogonality relation
\begin{align*}
&\int_{0}^{\infty}\int_{0}^{\pi}\int_{0}^{2\pi}r^{2\mu_1+2\mu_2+2\mu_3}\rvert\sin\theta\rvert^{2\mu_1+2\mu_2}\rvert\cos\theta\rvert^{2\mu_3}\rvert \cos \phi\rvert^{2\mu_1}\rvert \sin \phi\rvert^{2\mu_2}\;r^2\sin\theta \;\mathrm{d}r\,\mathrm{d}\theta\,\mathrm{d}\phi
\\
&R_{n_r'}(r)R_{n_r}(r)\Theta_{\ell'}^{(s_3')}(\theta)\Theta_{\ell}^{(s_3)}(\theta)\Phi_{m'}^{(s_1',s_2')}(\phi)\Phi_{m}^{(s_1,s_2)}(\phi)=\delta_{n_r,n_r'}\delta_{\ell,\ell'}\delta_{s_3,s_3'}\delta_{m,m'}\delta_{s_1,s_1'}\delta_{s_2,s_2'}.
\end{align*}
In analogy with the standard three-dimensional oscillator, the symmetries responsible for the separation of variables in spherical coordinates are related to the Dunkl ``rotation'' generators. Indeed, one has that the operator
\begin{align}
J_3^2=\left\{\frac{1}{i}\big(x_1 \mathcal{D}_2-x_2\mathcal{D}_1\big)\right\}^2=2\mathcal{B}_{\phi}+2\mu_1\mu_2(1-R_1R_{2}),
\end{align}
is diagonal on the separated wavefunction in spherical coordinates and has eigenvalues
\begin{align*}
J_3^{2}\;\Psi_{n_{r},\ell,m}^{(s_1,s_2,s_3)}(r,\theta,\phi)=\big[4m(m+\mu_1+\mu_2)+2\mu_1\mu_2(1-s_1s_2)\big]\;\Psi_{n_{r},\ell,m}^{(s_1,s_2,s_3)}(r,\theta,\phi).
\end{align*}
Furthermore, a direct computations shows that the Dunkl total angular momentum operator
\begin{align*}
&\mathbf{J}^2=
\left\{\frac{1}{i}\big(x_2 \mathcal{D}_3-x_3\mathcal{D}_2\big)\right\}^2+
\left\{\frac{1}{i}\big(x_3 \mathcal{D}_1-x_1\mathcal{D}_3\big)\right\}^2+
\left\{\frac{1}{i}\big(x_1 \mathcal{D}_2-x_2\mathcal{D}_1\big)\right\}^2
\end{align*}
has the following expression in spherical coordinates:
\begin{align*}
\mathbf{J}^2=
2&\left(\mathcal{N}_{\theta}+\frac{1}{\sin^2\theta}\mathcal{B}_{\phi}\right)+2\mu_1\mu_2(1-R_1R_2)+2\mu_2\mu_3(1-R_2R_3)+2\mu_1\mu_3(1-R_1R_3)
\\
&\quad+\mu_1(1-R_1)+\mu_2(1-R_2)+\mu_3(1-R_3),
\end{align*}
where $\mathcal{N}_{\theta}$ and $\mathcal{B}_{\phi}$ are as in \eqref{Azi} and \eqref{B}. It thus follows that the separated wavefunctions in spherical coordinates $\Psi_{n_r,\ell,m}^{(s_1,s_2,s_3)}(r,\theta,\phi)$ satisfy
\begin{align*}
\mathbf{J}^2\Psi_{n_r,\ell,m}^{(s_1,s_2,s_3)}(r,\theta,\phi)=\lambda_{\ell,m}\Psi_{n_r,\ell,m}^{(s_1,s_2,s_3)}(r,\theta,\phi),
\end{align*}
with
\begin{align*}
&\lambda_{\ell,m}=4(\ell+m)(\ell+m+\mu_1+\mu_2+\mu_3+1/2)+2\mu_1\mu_2(1-s_1s_2)+2\mu_1\mu_3(1-s_1s_3)
\\
&\quad +2\mu_2\mu_3(1-s_2s_2)+\mu_1(1-s_1)+\mu_2(1-s_2)+\mu_3(1-s_3).
\end{align*}
\section{Conclusion}
In this paper, we have examined the isotropic Dunkl oscillator model in three-dimensional Euclidean space and we have shown that this system is maximally superintegrable. The symmetries of the model were exhibited and the invariance algebra they generate, called the Schwinger-Dunkl algebra $sd(3)$, has been seen to be a deformation of the Lie algebra $\mathfrak{u}(3)$ by involutions. So far, we have examined Dunkl systems with oscillator type potentials. In view of the superintegrability and importance of the Coulomb problem, the examination of the Dunkl-Coulomb problem is of considerable interest. This will be the subject of a future publication.
\section*{Acknowledgements}
V.X.G. holds an Alexander-Graham-Bell fellowship from the Natural Sciences and Engineering Research Council of Canada (NSERC). The research of L.V. is supported in part by NSERC.
\section*{References}

\end{document}